\documentstyle[epsf]{mn}
 
\begin{document} 

\title[Rugate filters for OH suppression in $I$ and $J$ bands]{Rugate
  filters for OH-suppressed imaging at near-infrared wavelengths}
 
\author[A.R.Offer and J.Bland-Hawthorn]
 {Alison~R.~Offer$^1$ and Joss~Bland-Hawthorn$^2$\\
$^1$Anglo-Australian Telescope, Siding Spring Observatory, Coonabarabran, NSW 2357, Australia \\
$^2$Anglo-Australian Observatory, P.O. Box 296,  Epping, NSW 2121,  Australia }

\date{\today} \pagerange{\pageref{firstpage}--\pageref{lastpage}}
\pubyear{1997}

%%%%%%%%%%%%%%%%%%%%%%%%%%%%%%%%%%%%%%%%%%%%%%%%%%%%%%%%% 
 
\label{firstpage} 
 
\maketitle

\begin{abstract}
  Ground based observations at near-infrared wavelengths are
  severely affected by atmospheric OH bands. Many authors have
  recognized the  potential gains in sensitivity from suppressing
  these features. Dispersive instruments show some promise but are
  both expensive and complicated to build. OH suppression filters
  using single \cite{herbst} or periodic notches \cite{jones} have
  the advantage of simplicity but significant gains have not yet
  been realised.
 
  Rugate filters (with graded index inhomogeneous coatings) offer key
  advantages for astronomical imaging. It is possible to produce a
  transmission profile comprising a series of irregular and sharply
  defined bandpasses. We demonstrate through numerical
  simulation of rugate filters that it should be possible to achieve 95\%
  suppression of the OH features in the $J$ photometric band, while
  retaining roughly half of the spectral coverage. This would lead to
  extraordinary gains in sensitivity even for observations of
  continuum sources.  In addition, these filters 
  allow longer exposures before the detector
  saturates on the sky background.  $I$ and $z$-band filters
  can also be envisaged.
  
  In 1\arcsec\ seeing, a $J$-band rugate filter used in conjunction
  with a 4m telescope would detect a $J=23$ continuum source at
  5.0$\sigma$ in a single 10 min exposure.  In comparison, a
  conventional $J$ filter requires multiple exposures for a 10 minute
  integration time and achieves only a 2.5$\sigma$ detection. For
  emission line sources, the rugate filter has an even bigger
  advantage over conventional filters, with a fourfold increase in
  signal:noise ratio possible in certain instances.
  
  Astrophysical studies which could benefit from rugate filters are
  searches for very low mass stars and galaxy evolution out to $z=3$.
\end{abstract}

\section{Introduction} 
 
The night sky spectrum in the near infrared is dominated by bright
emission lines from vibrationally and rotationally excited OH
molecules in the mesosphere and lower thermosphere (80-100km
altitude). The principle source of the OH molecules is believed to
be the reaction between atomic hydrogen and ozone \cite{chemistry}
which produces OH in vibrational states up to $\nu=9$. The
subsequent decay of the excited OH, along with a small
contribution from other processes, is believed to be responsible
for the OH emission \cite{bates,letexier}. The OH emission can
vary both temporally and spatially across the sky
\cite{johnston,ramsay}.  In particular, the propagation of density
and temperature differences through the atmosphere can cause large
variations in both the relative and absolute intensity of the OH
lines with periods as short as a few minutes.  The presence of
this intense and variable background is a serious problem for
earth-based astronomical observations in the near infra-red. In
the $J$ photometric band (1.12\hbox{$\mu\rm m$} to
1.38\hbox{$\mu\rm m$}) around 95\% of the sky background is in the
form of line emission, principally OH.  The OH emission in the $I$
photometric band is less intense but is still a large percentage
of the background. Clearly there would be large potential gains if
the OH emission could be suppressed, both in terms of the
improvement in signal to noise ratio if the sky background were
reduced by over 90\%, and in terms of the increased exposure time
a much reduced and more stable sky background would allow.

There has been an increasing interest in methods of suppressing the OH
background. These can be grouped into dispersive solutions
\cite{maihara1,iwamuro,content1,content2,piche} and filter based
solutions \cite{jones}. Filter based solutions have the advantage of
being relatively simple, especially for imaging purposes. However,
numerical simulations of filter based solutions have been generally
disappointing \cite{herbst,jones} and attention has been directed
towards dispersive solutions.  Filter based solutions have suffered
from the inflexibility of simple filter profiles.  Herbst
\shortcite{herbst} investigated bandpass filters but found only a
moderate improvement in the sensitivity for the $J$ band. Jones et al. 
\shortcite{jones} investigated the use of multiple band-pass filters
but the improvement in sensitivity was limited by problems matching
the periodic transmittance profile to the OH emission.  However,
recent advances in filter technology, driven by the development of
highly wavelength specific laser rejection filters \cite{johnson},
mean that filters with almost any transmittance profile can be
designed and tailored to a particular problem. It is therefore
possible to design of an on-off filter with pass bands at any desired
wavelength.  A filter blocking most of the OH lines individually is
infeasible but there are no technical limitations to building an
on-off filter which masks regions of the spectrum overwhelmed by
bright OH vibrational bands, whilst retaining roughly half the
spectral coverage. Such rugate devices can now be manufactured and
promise major gains for wide-field imaging.

In this paper, it is demonstrated that significant gains in the
broad-band signal to noise ratio can be envisaged using the ideas of
rugate filter design.  By exploiting the fact that the intense OH
lines at wavelengths of less than 1.4\hbox{$\mu\rm m$} are grouped into distinct
and non-overlapping bands, large increases in the signal to noise
ratio can be achieved in the $J$ photometric band. The gains in the
$I$ photometric band depend on the assumed background continuum level,
but worthwhile increases in the signal to noise ratio can be found if
the continuum is small. The first section of the paper introduces the
rugate filter concept.  A basic design strategy is outlined in section
2 and section 3 describes the signal to noise ratio calculations. The
results are presented in section 4. Section 5 discusses the
practicality of manufacturing such filters and a brief discussion on
how the concepts used could be extended and applied to other problems.
The paper concludes with a summary in section 6.

\section{Rugate filters} 
 
Rugate\footnote{The name derives from the Latin word for `wrinkled'
  which describes the refractive index structure.} filters are thin
film filters with refractive indices that vary continually through the
thin film coating \cite{johnson}. They are manufactured by carefully
controlling the chemical composition of a mixture of thin film
materials with different individual refractive indices during
deposition. In the simplest case, a rugate filter with an infinitely
long sinusoidal refractive index profile gives a transmittance profile
with a single minimum at a wavelength of $\lambda = 2n_a x_P$ where
$n_a$ is the average refractive index and $x_P$ is the geometric
period. The width of the minimum is related to the amplitude of the
sinusoid (a narrow minimum corresponds to a small amplitude). For a
given period and amplitude, the sharpness of the transmittance profile
is related to the number of periods included in the filter (a longer
filter gives a sharper transmittance minimum). A superposition of two
sinusoids gives two transmittance minima with positions dependent on
the individual periods and widths dependent on the individual
amplitudes.
 
The relationship between a sinusoidal filter and a quarter wave stack
can by understood by analogy with Fourier theory. A quarter wave stack
has a square refractive index profile, and therefore requires an
infinite number of sinusoids to represent it, with the fundamental
geometric period being $P = \lambda/2n_a$.  The quarter wave stack
thus gives transmittance minima at $\lambda = 2n_a x_P/m$
$(m=1,2,3,...)$.

Between the two limits there is an infinite range of possible filter
profiles. In principle filters giving complex and sharply varying
transmittance curves can be designed by tailoring the refractive index
profiles.  Apodisation of the profiles, and quintic matching layers at
the air/rugate and rugate/substrate interfaces can be used to reduce
reflection losses and sidelobes in the transmission profiles of real
filters with a finite extent \cite{apodise,quintics1,quintics2}.

The possibility of being able to design filters with a small number of
transmittance minima at any chosen wavelength and for any desired
width has not previously being considered when considering filter
based solutions to the OH suppression problem.
 
\section{Filter Design} 

The design of a filter with a given transmittance profile is a typical
example of an inverse problem. Although the relationship between the
transmission profile and refractive index profile can be understood by
analogy to Fourier theory, no exact transformation between the two
exists. The forward problem -- given a refractive index profile, find
its transmittance profile -- is straight forward, but the inverse of
this is, in general, non-trivial and the solution is non-unique. There
is a wide range of literature dealing with solutions to the inverse
problem using either methods based on Fourier transforms or
optimisation techniques, or a combination of the two
\cite{opticsrefs}.  The details and relative merits of the methods are
beyond the scope of this paper. To illustrate the idea of using rugate
filters for OH suppression we only require a basic method to optimise
a profile. The aim here is not to design a perfect filter but to
illustrate that realistic transmission profiles can be produced that
significantly reduce the OH background in the near infra-red. To this
end, it was assumed that the refractive index profile could be written
as the sum of a small number of sinusoids and the parameters were
optimised using a genetic algorithm (GA) approach \cite{gabook}.  The
use of Genetic Algorithms to design multilayer filters has been
discussed by Martin et al.  \shortcite{ga_opt} and Greiner
\shortcite{greiner}. Martin et al.  and Greiner both use GA to optimise
the thicknesses and refractive indices of the individual layers in
multilayer stacks.  The implementation of GA used here was slightly
different. The GA sought to optimise the amplitude, phase and period
of a sum of sinusoids representing the refractive index profile. The
algorithm used was based on the method outlined by Charbonneau
\shortcite{ga}. It is described in more detail in the appendix but in
outline it entails constructing an initial random population of
solutions.  The members of the population are ranked according to
their fitness (or the extent to which they minimise or maximise some
function). The population is then allowed to evolve towards an optimal
solution through a series of breeding and mutation operations. This
approach had the advantage of allowing the possiblility of 
simultaneous optimisation of the refractive indices through the 
filter and the transmission profile itself. It was therefore possible to 
seek directly the transmission profile that gave the optimum improvement 
in sensitivity.

\begin{figure}
%\label{f-ohlines} 
 \ \centering{ \epsfxsize=3.3in \epsfbox{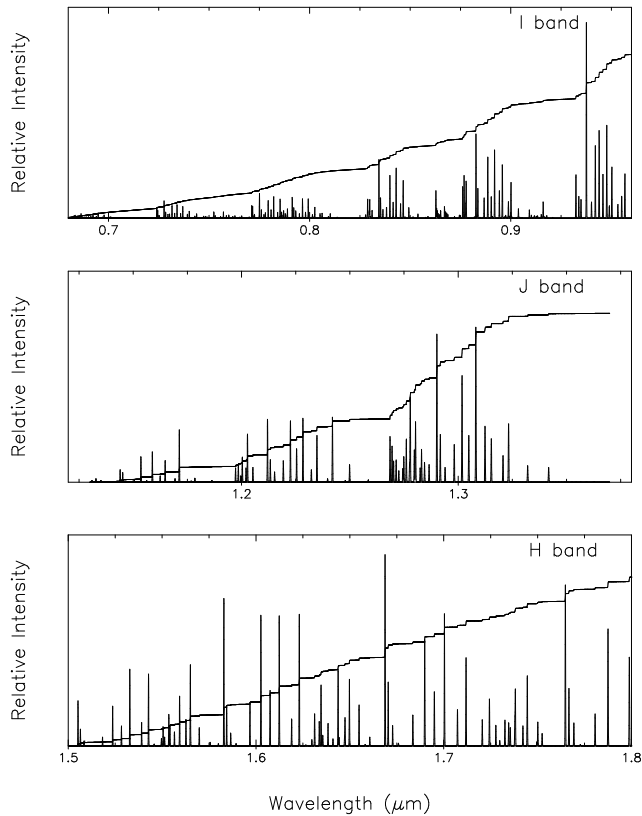} }
\caption{The sky spectrum in the $I$, $J$ and $H$ bands. Note that the $J$ band
  OH lines are grouped into a series of non-overlapping vibrational
  bands.  In the $I$ band there is some overlap but the most intense
  OH emission is concentrated in a few spectral regions.  In the $H$
  band the vibrational bands overlap and the OH lines are more evenly
  distributed. The continuous line is the cumulative background intensity
  as a function of wavelength through the band.}
\end{figure}  

The sky spectrum in the $I$, $J$ and $H$ bands is shown in figure~1.
The emission lines are principally OH although there are oxygen bands
around 1.27\hbox{$\mu\rm m$}\ and 0.865\hbox{$\mu\rm m$}. Also shown is the cumulative
background intensity as a function of wavelength.  In the $I$ and $J$
bands the OH lines are grouped into a few main bands separated by
relatively clean windows and the cumulative background intensity has
a step-like profile.  This suggests the approach of seeking a $I$ and
$J$ filters that give a transmittance close to unity in the windows
and close to zero around the OH bands. An alternative approach would
be to attempt to block individual OH lines. The latter approach is the
method adopted by most dispersive OH suppression methods. It could
potentially give a greater improvement in the signal to noise ratio
for continuum objects because more object flux would be retained.
However, because of the intrinsic width of filter profiles, this is
infeasible for filter based solutions and the first approach was
adopted. Note that in the $H$ band the OH vibrational bands overlap
and the cumulative intensity has a smoother gradient. Effective OH
suppression would require a far more complicated transmission profile
so consideration was limited to the $I$ and $J$ bands.

In outline, the basis approach used for this illustration was as
follows:
\begin{enumerate}
\item{ Seek for an `ideal' filter profile with a few blocking bands
    optimally positioned for maximum signal:noise improvement}
\item{ Use the `ideal' filter as an initial goal for a genetic
    algorithm seeking to optimise a small number of sinusoids. This
    produces an approximation to a relatively simple square wave
    profile.}
\item{ As the `ideal' transmittance profile is approached, optionally
    switch to a genetic algorithm seeking to maximise a function of
    the percentage of object flux retained and the increase in the
    signal:noise ratio. This step improves the theoretical performance
    of the filter at the expense of producing more complex (therefore
    less robust) transmission profiles.}
\end{enumerate}

To select an `ideal' filter, it was assumed that the maximum
signal:noise gain would be achieved when the boundaries of the block
were set just outside two OH lines. The blocks were positioned
iteratively by summing the source and background flux between pairs of
OH lines and positioning the blocks sequentially such that the
signal:noise improvement was maximised.
 
To calculate a rugate filter a genetic algorithm approach was used to
optimise a refractive index profile of the form:
\begin{equation} 
  n(x<0) = 1
\end{equation} 
\begin{equation} 
\label{eq_riform} 
n(0<x<x_0) = n_a + \sum_{i=1}^m \delta_i \sin(\alpha_i x + \phi_i)
\end{equation} 
\begin{equation} 
  n(x>x_0) = n_s
\end{equation} 
where $n_s$ is the assumed refractive index of the substrate, $n_a$ is
the assumed average refractive index of the rugate film, $m$ is the
number of sinusoids included and $x_0$ is the thickness of the film.
The parameters, $\alpha_i, \delta_i$ and $\phi_i$ were optimised using
genetic algorithms.
 
Initially values of $\alpha_i, \delta_i$ and $\phi_i$ were sought to
maximise the function:
\begin{equation}
\label{eq_fit1}
{F} = \left({1 - {\sum_{i=1}^{n_k} (T_{\rm filt}(k_i)
      -T_{\rm ideal}(k_i))^2\over n_{k}}}\right)
\end{equation}
where $T_{\rm filt}(k)$ is the calculated transmittance of the filter at a
wavenumber $k$ and $T_{\rm ideal}(k)$ is the predetermined `ideal'
transmittance profile. As the function began to converge the strategy
was switched to seeking explicitly parameters that maximised some
function of the relative improvement in the broad-band signal:noise
ratio, $G$. To reflect the balance between the improvement in signal
to noise ratio and the percentage of object flux retained ($\tau$) an
function of the form:
\begin{equation}
\label{eq_fit2}
{F} = G^m\tau^n
\end{equation}
was used where $m$ and $n$ are positive powers.
 
For a given refractive index profile the transmittance was calculated
by dividing wavenumber space up into a series of discrete bins of
width $\delta k\approx 2\times 10^{-3}\mu{\rm m}^{-1}$ and approximating
the continually varying refractive index profile by a series of thin
homogeneous layers each with a constant refractive index. The
transmittance at each value of $k$ can then be calculated by the
matrix method \cite{levi}.  The transmittance profiles of the filters
were calculated assuming a collimated beam at normal incidence.
Throughout it was assumed that the filter materials were
non-dispersive and absorption was assumed to be negligible.

The genetic algorithm approach is excellent at locating the region of
the global minima (or maxima) in parameter space , but the convergence
to the exact minima (or maxima) is very slow and other optimisation
techniques should be used to refine the solution \cite{gabook,ga_opt}.
A simple iterative optimisation technique was used to fine tune the
solution some of the filter profiles. The filter was first converted
from a series of layers with a stepped refractive index profile to an
equivalent multilayer stack of high/low refractive index pairs.
Southwell \shortcite{southwell} has shown that for thin layers
($nt<<\lambda$ where $n$ is the refractive index and $t$ the
thickness), a single layer of intermediate refractive index can be
replaced by two equivalent layers of high and low refractive indices
with thickness given by:
\begin{equation}
  t_{n_H} = t\left({ 1 + {n_H^2 -n^2\over n^2 - n_L^2}}\right)^{-1}
\end{equation}
\begin{equation}
  t_{n_L} = t - t_{n_H}
\end{equation}
By alternating the order of the equivalent layers ($(n_H n_L)(n_L n_H)
(n_H n_L)$...) this can be done without increasing the number of
layers, provided the thin layer assumption holds. Provided the layers
are sufficiently thin the new filter has the same transmittance
properties as the original rugate filter. The resulting multilayer
stack was optimised by choosing an initial step size, $\delta t$ and
looping over the layers, incrementing the thicknesses of the
individual layers by $\pm \delta t$. For each layer the new fitness
was calculated and if the change in thickness increased the fitness
the change was retained. Once all the layers had been tested, if no
change had been made to the filter the step size was decreased.
Otherwise it was increased. The characteristic matrix of the filter is
a product of the characteristic matrices of the individual layers.
This property means that a simple optimisation scheme that
sequentially tests and changes each layer can be implemented
efficiently with the characteristic matrix at each step being given
by:
\begin{equation}
  {\bf C}_{filter}(t_i+\delta t) = {\bf C}_{1\to i-1} {\bf
    C}_{i}(t_i+\delta t) {\bf C}_{i}^{-1}(t_i) {\bf C}_{i\to N}
\end{equation}
where ${\bf C}_i$ is the characteristic matrices of the $i$th layer
and:
\begin{equation}
  {\bf C}_{k\to j} = \left({\Pi_{l=k}^{j} {\bf C}_{l}(t_l)}\right)
\end{equation}

Note that after the secondary optimisation, the filters are no longer
true rugate filters in the sense that they no longer have continually
varying refractive index profiles. However, they are derived directly
from rugate filters.

\section{Signal:noise improvement \label{sec-snr}} 
 
The calculation of the signal:noise improvement followed the method
given by Jones et al.  \shortcite{jones}. The SNR without a filter is
assumed to be:
\begin{equation} 
\label{eq_snr0} 
(S:N)_0 = {\epsilon S t \over (\epsilon t (S + B) + p(Dt +
  \sigma_R^2))^{1/2}}
\end{equation} 
where $S$ is the integrated source flux, $B$ is the integrated
background flux (continuum plus line emission) both in electrons
s$^{-1}$ summed over the $p$ pixels, $D$ is the dark current ($\rm
e^-\hbox{pix}^{-1} s^{-1}$), $\sigma_R$ is the
rms detector read noise (e$^-$rms), $p$ is the number of pixels
illuminated by the source, $\epsilon$ is the efficiency of the
detector system and $t$ is the exposure time in seconds. 
Following Jones et al.  \shortcite{jones}, the introduction
of the filter reduces the object flux to $\tau S$ and the
background flux to $\beta B$. The SNR becomes:
\begin{equation} 
\label{eq_snrf} 
(S:N)_F = {\epsilon \tau S t \over (\epsilon t (\tau S + \beta B) +
  p(Dt + \sigma_R^2))^{1/2}}
\end{equation} 
The improvement (or otherwise) in the signal:noise ratio is given by:
\begin{equation} 
\label{eq_gain} 
{(S:N)_F\over (S:N)_0} = \tau \left({\gamma + 1 + k(t,B) \over
    \tau\gamma + \beta + k(t,B)}\right)^{1/2}
\end{equation} 
where $k(t,B)$ is a function of $t$ and $B$ that is independent of the
filter parameters and $\gamma=S/B$. The assumed instrument parameters
are listed in Table~\ref{t-instr}. For the $J$ band a constant quantum
efficiency of 70\% was assumed with a read noise of 30e$^-$ and dark
current of 5e$^-$s$^{-1}$. The assumed read noise in $J$ is conservative
and modern CCD's can have a read noise as low as 5e$^-$ but the values
were chosen to be consistent with previous calculations \cite{jones}. For the $I$
band the quantum efficiency curve of an MITLL-CCD deep depletion
device was assumed. This gives a QE of around 90\% at wavelengths less
than 0.85\hbox{$\mu\rm m$}\ dropping to around 60\% at 0.95\hbox{$\mu\rm m$}. A read noise
of 1e$^-$ and negligible dark current were assumed.
 
\begin{table} 
  \centering
\caption{Assumed Instrumental Parameters} 
\label{t-instr} 
\begin{tabular}{lc} 
  \hline
  Telescope Diameter  & 3.9m \\
  System Efficiency, $\epsilon$ &  50\%   \\
  Image Scale & 0.5\arcsec/pixel \\
  Detector Read Noise  &   J: 30e$^-$rms  \\
  &   I: 1e$^-$rms\\
  Dark Current 
  & J: 5$\rm e^-\hbox{pix}^{-1} s^{-1}$ \\
  & I: 0$\rm e^-\hbox{pix}^{-1} s^{-1}$ \\
  Quantum Efficiency & J: 70\% \\
  & I:90\% @$<0.85$\hbox{$\mu\rm m$}, 60\% $@ 0.95$\hbox{$\mu\rm m$}\\
  size of source & 0.5\arcsec, 1\arcsec\ , 2\arcsec\ \\
  \hline
\end{tabular} 
\end{table}

In order to keep the calculations as general as possible
continuum sources were used throughout. For the $J$ band a
continuum source with a magnitude of 22 was assumed. The flux was
calculated using the zero magnitude values of Koorneef
\shortcite{koorneef}.  This gave a total flux of 47 photons s$^{-1}$
m$^{-2}$ arcsec$^{-2}$ band$^{-1}$ which was assumed to be constant across
the band. 
 
It is also necessary to estimate the background flux across the band. Maihara
et al.  \shortcite{maihara2} quote a night sky continuum of about 560
photons s$^{-1}$ m$^{-2}$ arcsec$^{-2}$ \hbox{$\mu\rm m$}$^{-1}$\ measured at
1.65 \hbox{$\mu\rm m$}.  The continuum is expected to be less at shorter
wavelengths and a night sky continuum of 300 photons s$^{-1}$ m$^{-2}$
arcsec$^{-2}$ \hbox{$\mu\rm m$}$^{-1}$ was assumed \cite{sobolev}. This gave a
$J$ band total of 72 photons s$^{-1}$ m$^{-2}$ arcsec$^{-2}$ band$^{-1}$.  The
$J$ band emission line data used by Jones et al.  \shortcite{jones} was
used \cite{piche2}. The emission lines were represented by Gaussians of width
1\AA\ centred at the line locations, such that the integrated flux was equal to
the line intensity.  The total flux from emission lines in the $J$ band was 3560
photons s$^{-1}$ m$^{-2}$ arcsec$^{-2}$ band$^{-1}$.

For the $I$ band calculation an $I$ band centred at 0.84\hbox{$\mu\rm m$}\ 
with a FWHM of 0.239\hbox{$\mu\rm m$}\ and a zero magnitude flux
of $9.9\times 10^{-10}$ erg cm$^{-2}$ s$^{-1}$ \AA$^{-1}$ was
assumed. This extends further into the red than many conventional $I$
filters and covers a substantial portion of the spectrum included in
$z$ filters  (see for example Fukugita et~al.
\shortcite{fuku}). A continuum source of magnitude 24 was used. The
night sky in the $I$ band is much less intense than in the $J$ band,
with a magnitude  $>19$ arcsec$^{-2}$ compared with
about 15 arcsec$^{-2}$ for the $J$ band. In particular the OH
emission is much weaker. The vibrational transitions giving rise to
the $I$ band OH emission change the OH vibrational quantum number,
$\nu$, by $\Delta\nu=4$ and $\Delta\nu=5$. The $J$ band OH emission
is from $\Delta\nu=3$ transitions. The transition probability falls
of rapidly with increasing $\Delta\nu$ \cite{bates} with the
consequence that the $I$ band OH lines are much fainter. For example,
Harrison and Kendall \shortcite{harrison} measured a total intensity
for the seven principle OH bands between 0.72 and 0.96\hbox{$\mu\rm
m$}\ that was roughly one tenth of the intensity measured for the
three principle OH bands in the $J$ photometric band. Llewellyn
\shortcite{llewellyn} predicted a similar intensity distribution.

Estimation of the intensity of the OH lines is one of the major
uncertainties in all these calculations. The intensities are highly
variable and poorly known in all wavelength bands.  The positions and
relative intensities of the $I$ band OH lines at wavelengths less than
0.94\hbox{$\mu\rm m$} were taken from a fit to a high resolution spectrum kindly
obtained for us by C.G. Tinney with the CASPEC echelle at the ESO 3.6m
telescope. The data were obtained in a 6900 sec exposure at 9 km
s$^{-1}$ resolution during dark time.  The absolute intensities were
scaled so that the total intensity of the OH bands and the
0.865\hbox{$\mu\rm m$} oxygen band equalled the total intensity measured by
Johnston \& Broadfoot \shortcite{johnston} at an altitude of 2750m and
latitude of $32^\circ26.6'$N. The band intensities measured by
Johnston and Broadfoot \shortcite{johnston} are approximately half the
band intensities measured by Harrison \& Kendall \shortcite{harrison}
but are similar to, or slightly larger than, the predicted intensities
calculated by Llewellyn et al.  \shortcite{llewellyn}.  The line positions
of the band around 0.945\hbox{$\mu\rm m$}\ were taken from the spectra of 
Abrams et al.  \shortcite{abrams} and Osterbrock et al. 
\shortcite{osterbrock}
and the relative intensities were estimated from the spectra of
Osterbrock et al.  \shortcite{osterbrock} and Steed \& Baker \shortcite{steed}.
The lines were scaled so that the total intensity was equal to that 
calculated by Llewellyn et al.  \shortcite{llewellyn}. This gave a total 
of 152 line photons s$^{-1}$ m$^{-1}$ arcsec$^{-2}$ between 0.65 and 0.96\hbox{$\mu\rm m$}.

Since the $I$ band OH lines are weaker than in $J$, the value of the
background continuum is more critical in the signal to noise ratio
calculations, but it's value is equally as uncertain as the $J$ band
value.  Noxon \shortcite{noxon} measured 130 photons s$^{-1}$ m$^{-2}$
arcsec$^{-2}$\hbox{$\mu\rm m$}$^{-1}$ at 0.85\hbox{$\mu\rm m$} in a dark sky and this was the
value adopted for the calculations.  It was assumed that the continuum
was flat, giving 40 continuum photons s$^{-1}$ m$^{-1}$ arcsec$^{-2}$ in the
same wavelength range. The resulting $I$ band sky spectrum gave a
magnitude of 19.9 arcsec$^{-2}$ which is consistent with dark sky
measurements in the $I$ band.

For both the $I$ and $J$ bands, the total source and background fluxes
were multiplied by the atmospheric transmission as a function of
wavenumber and binned into a discrete grid in wavenumber space.

The improvement in signal to noise ratio has
been calculated assuming equal exposure times with and without the
filter. If the exposure time is limited by sky
saturation, the maximum improvement in signal to noise ratio  for a
single exposure  is potentially much higher. If $t$ is the time
taken before saturation without the filter, then the exposure times
with the filter can be increase to $t'=t/\beta$. The resulting
improvement in S:N for a single exposure is:

\begin{equation} 
\label{eq_gain2} 
{(S:N)_F(t/\beta)\over (S:N)_0(t)} \approx {1\over \beta^{1/2}}
{(S:N)_F(t)\over (S:N)_0(t)}
\end{equation} 

\section{Calculations and results} 
  
\subsection{J-band Filter}

The OH lines in the $J$ band are shown in figure~1.  The bands fall
into three distinct regions so an `ideal' blocking filter with three
blocking bands was chosen.  The `ideal' blocking filter is shown in
figure~2. This filter increases the broad-band signal to noise ratio
for a 1\arcsec\ magnitude 22 source by a factor of 2.  Higher gains are
possible with more and finer blocking lines.  However, for the sake of
illustration, we chose to keep the filter relatively simple at this
stage.

\begin{figure} 
\label{f-jtrans} 
  \ \centering{ \epsfxsize=3.3in  \epsfbox{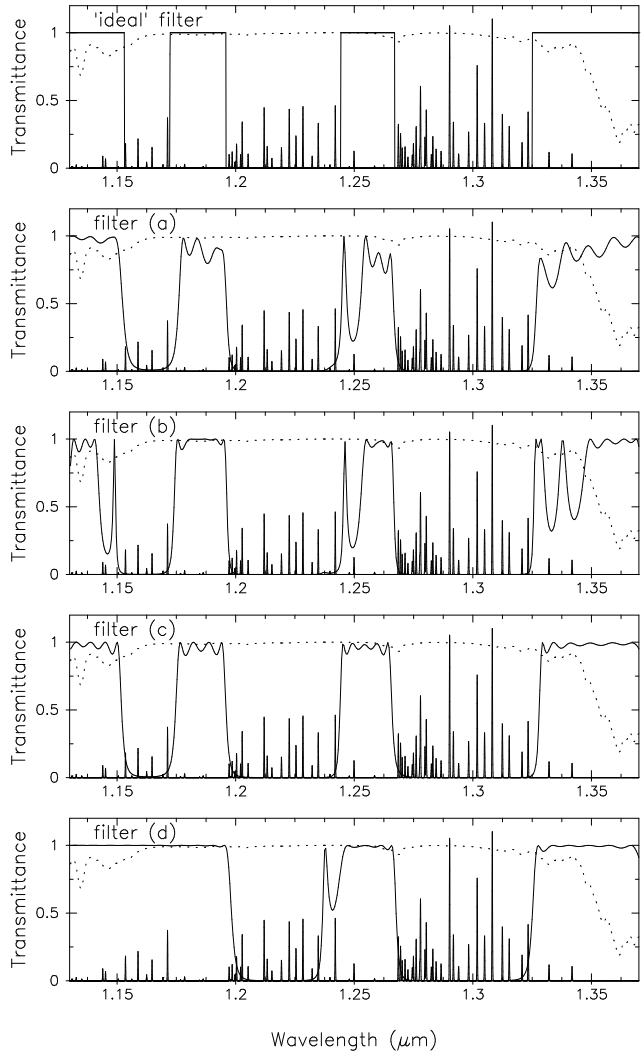} }
\caption{The transmittance profiles as a function of wavelength for the `ideal'
  filter and filters (a) to (d) described in the text. The positions
  of the emission lines and the atmospheric transmittance profile are
  also shown.}
\end{figure}

Using the `ideal' filter as an initial template, four filters were
derived whose transmittance profiles are also shown in figure~2.  The
transmittance calculations assumed a collimated beam at normal
incidence. The properties of these filters are summarised in table~2.
All filters assumed a substrate refractive index of $1.47$. Filter~(a)
is a true rugate filter designed by GA alone. The RI profile is the
sum of 10 sinusoids constrained to lie between $\alpha=4\pi
n_a/(1.39\mu m)$ and $\alpha=4\pi n_a/(1.12\mu m)$ with no constraint
on the minimum and maximum allowed refractive index. It was designed
to minimise the function $G^4\tau^{1/4}$ where $G$ is the factor by
which the signal to noise ratio is increased. The narrow dips in the
transmittance corresponding with the outlying emission lines are a
feature of the GA optimisation. The resulting refractive index profile
is shown in figure~2. It covers a range of 1.34 to 1.66. With a small
amount of clipping which does not degrade the transmittance profile,
it is within the range allowed by real materials (eg 1.35 (cryolite)
and 2.3 (ZnS)).
 
\begin{figure}
  \ \centering { \epsfxsize=3.3in  \epsfbox{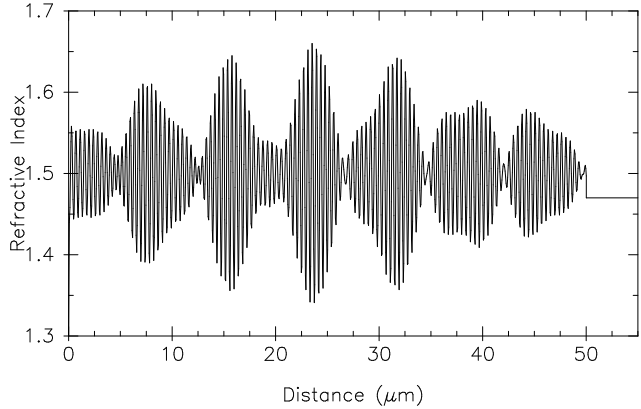} }
%\label{f-jriprofile}
\caption{ The Refractive Index Profile for Filter (a)}
\end{figure}

\begin{table*} 
\begin{minipage} {140mm} 
\caption{Summary of Filter Properties} 
\label{t-filtprop} 
\begin{tabular}{lcccccc} 
  \hline Filter & Description & Average RI & Min RI & Max RI &
  Thickness & No. of Layers
  \\
  Ja & True rugate maximising $G^4\tau^{1/4}$ & 1.5 & 1.34 & 1.66 &
  50\hbox{$\mu\rm m$} &
  \\
  Jb & Optimised to maximise $G^4\tau$ & 1.5 & 1.35   & 1.62 & 50\hbox{$\mu\rm m$} & 500 \\
  Jc & Optimised to `ideal' filter  & 1.5 & 1.35 & 1.62 & 50\hbox{$\mu\rm m$} & 500 \\
  Jd & Optimised to maximise $G\tau$ & 1.5 & 1.35 & 1.62 & 50\hbox{$\mu\rm m$} & 500 \\
  \\
  I & Optimised to `ideal' filter  & 1.5 & 1.35 & 1.62 &
  50\hbox{$\mu\rm m$} & 500
  \\
  \hline
\end{tabular} 
\end{minipage}
\end{table*} 
 
Filters (b) and (c) resulted from further optimisation of the GA
solution shown in figure~2. The optimisation technique described in
section~3 converts the sinusoidal profiles into multilayer stacks so
the resulting filters are no longer true rugate filters although they
derive directly from them. The optimisation process is illustrated in
figure~3. The first panel shows the clipped 50\hbox{$\mu\rm m$} thick rugate
filter from which filters (b) and (c) were derived. For illustration,
a minimum and maximum refractive index of 1.35 and 1.62 was imposed
(eg. cryolite, alumina). The materials were assumed to be
non-dispersive and absorption was assumed to be negligible.  The
sinusoidal profile was approximated by a series of 0.1\hbox{$\mu\rm m$} thick
homogeneous layers shown in part in the second panel.  These in turn
were translated into a stack of high/low refractive index layers
(third panel). The final panel shows the same section of filter (b)
after optimisation. Filter (b) was optimised to maximise the function
$G^4\tau$ and filter (c) was optimised directly to fit the `ideal'
transmittance profile.

\begin{figure}
  \ \centering { \epsfxsize=3.3in \epsfbox{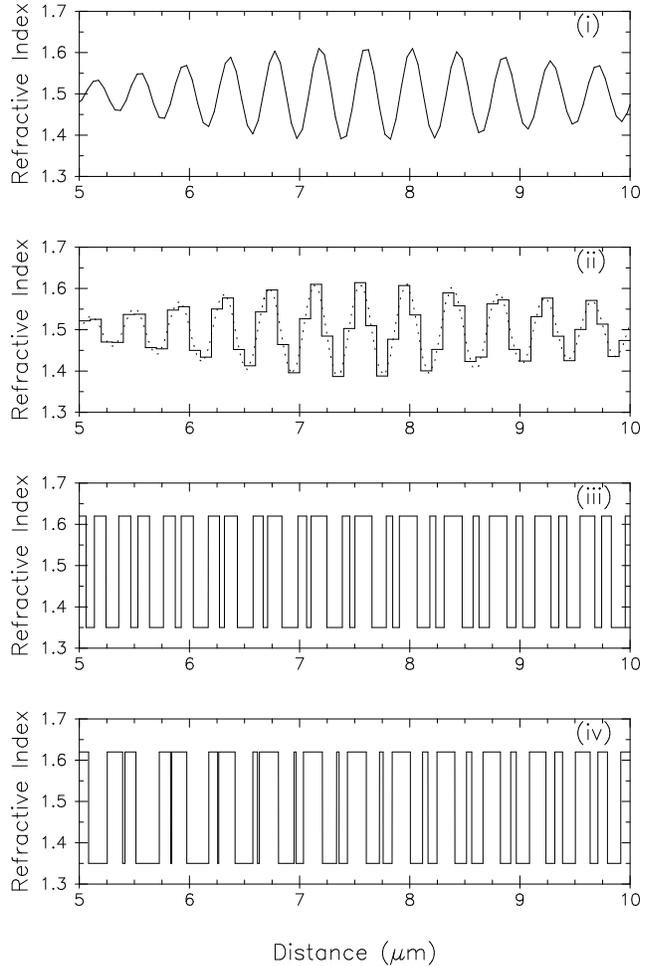} }
%\label{f-optim}
\caption{ Steps in the further optimisation of the GA results: (i) an
  enlarged section ($x=5\to 10$\hbox{$\mu\rm m$}) of the clipped version of the
  refractive index profile of filter(a) from which filters (b) and (c)
  were derived; (ii) an approximation to (i) with 0.1\hbox{$\mu\rm m$}
  homogeneous layers; (iii) conversion of (ii) into high/low
  refractive index stack; (iv) the result of optimisation (filter b).}
\end{figure}

Filter (d) in contrast was chosen to illustrate the balance between
spectral coverage and improved sensitivity. It was produced by
matching to an `ideal' transmittance profile with only two blocking
bands.

The results of the signal to noise calculations for the four filters
are summarised in table~3 and illustrated in figures~5 and 6.  Table~3
gives the detailed results for a magnitude 22 continuum object.
Figure~5 shows the calculated signal to noise ratio for filters (a) to
(d) and for no OH suppression (solid line) as a function of object
magnitude assuming a 1\arcsec\ continuum source. Figure~6 illustrates
the increase in exposure time before the detector saturates assuming a
1\arcsec\ magnitude 22 object and pixel saturation at $10^5$ electrons.

\begin{table*} 
\begin{minipage}{140mm}
  \centering
\caption{Improvement in Signal to Noise Ratio for a 600s 
  integration time. Results are given for objects of magnitude $J=22$
  and $I=24$.}
\label{t-snr} 
\vspace{0.3cm}
\begin{tabular}{lccccccccc} 
  \hline
  &   &  $G$  &   & $\tau$ &  $ \beta^{(a)}$ \\
  source size:&   2\arcsec\   & {\bf 1\arcsec}   & 0.5\arcsec\  \\
  \hline
  `Ideal' J   &  2.04 & \bf 2.03 &   1.98   &  0.47 & 0.050 \\
 \\
  Filter (Ja) &  1.74 & \bf 1.73 &   1.69   &  0.40 & 0.049 \\
  Filter (Jb) &  2.05 & \bf 2.03 &   1.97   &  0.36 & 0.030 \\
  Filter (Jc) &  1.76 & \bf 1.75 &   1.72   &  0.43 & 0.058 \\
  Filter (Jd) &  1.46 & \bf 1.46 &   1.45   &  0.60 & 0.166 \\
  \\
  `Ideal' I   & 1.23 & \bf 1.21 &   1.15    &  0.45 & 0.135 \\
 \\ 
  Filter (I)  &  1.17 & \bf 1.16 &   1.12   &  0.50 & 0.179 \\
  \hline
\end{tabular} \\
{$^a$ $1/\beta$ is approximately the factor by which the maximum exposure \\
  time before saturation is increased in the sky dominated limit}
\end{minipage}
\end{table*} 

\begin{figure} 
  \ \centering { \epsfxsize=3.3in \epsfbox{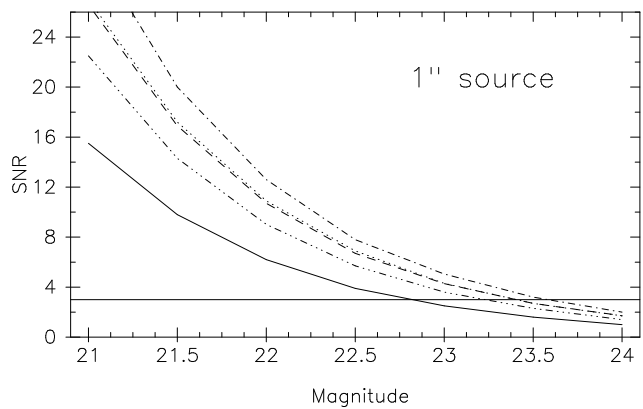} }
\caption{The signal to noise ratio as a function of magnitude for a
  1\arcsec\ with no OH suppression (solid line) and with the four
  filters (dashed line:filter (a); dash-dot: filter (b); dotted line:
  filter (c) and dash-dot-dot-dot: filter (d)). }
\label{f-jsnr} 
\end{figure}

\begin{figure} 
  \ \centering { \epsfxsize=3.3in \epsfbox{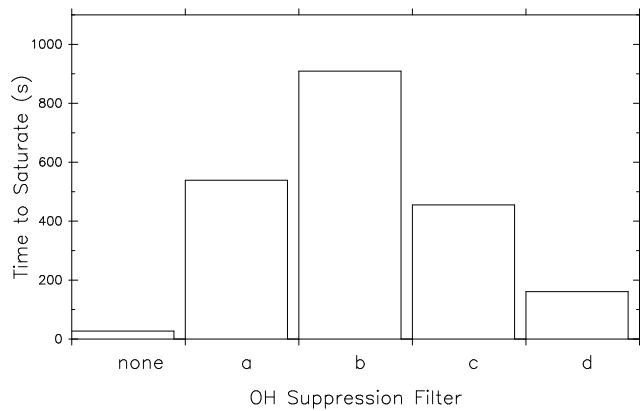} }
\caption{The time until saturation (assuming saturation at $10^5$ 
  electrons/pixel) for no OH suppression and for the four $J$ band OH
  suppression filters discussed.  }
\label{f-jtimes} 
\end{figure}

It can be seen that all rugate filters offer significant improvements
over standard photometric bands in terms of the increase in signal to
noise ratio for faint sources, with a factor of two improvement in
signal to noise ratio for filter~(b).  Improvements in signal to noise
ratio are possible for sources fainter than about magnitude 15 for a
2\arcsec\ source, 16.5 for a 1\arcsec\ source and 18 for a 0.5\arcsec
source.  A secondary advantage is the increase in exposure times
allowed, with maximum exposure times increasing by a factor of 30 for
filter (b).

Filter (b), optimised to a function of the form $G^n\tau^m$, gives the
best performance in terms of both the increase in signal to noise
ratio and the increased exposure time allowed but this is at the
expense of the spectral coverage. Filter (c) retains a greater
percentage of the object flux, at 44\% and still gives a significant
improvement in the signal to noise ratio. The transmittance profile of
filter (c) is more feasible in practice than filters (a) and (b).
Broad features will be less affected by small deviations in in the
refractive index profile and departure from normal incidence than
narrow dips aligned with individual OH lines.

Filter (d) gives a far better spectral coverage than the other three
filters with 60\% of the object flux being retained. This is at the
expense of some loss in sensitivity with a signal to noise improvement
of 1.45 for a 1\arcsec\ magnitude 22 continuum object. However, the
relatively large fraction of line emission retained limits the
increase in maximum exposure time and implies that the filter
performance will be far more dependent on the intensity of the OH
lines, which is a highly variable and uncertain quantity.

\subsection {I band filter}

The OH lines in the $I$ band were shown in figure~1. The OH bands at
the long wavelength end of the band are well separated, but the
windows between the bands are no longer so clear. In particular, there
is a relatively intense oxygen band around 0.865\hbox{$\mu\rm m$}. At shorter
wavelengths the bands are less intense but overlap so again the
windows between bands are not so clean.

\begin{figure} 
%\label{f-itrans} 
  \ \centering { \epsfxsize=3.3in \epsfbox{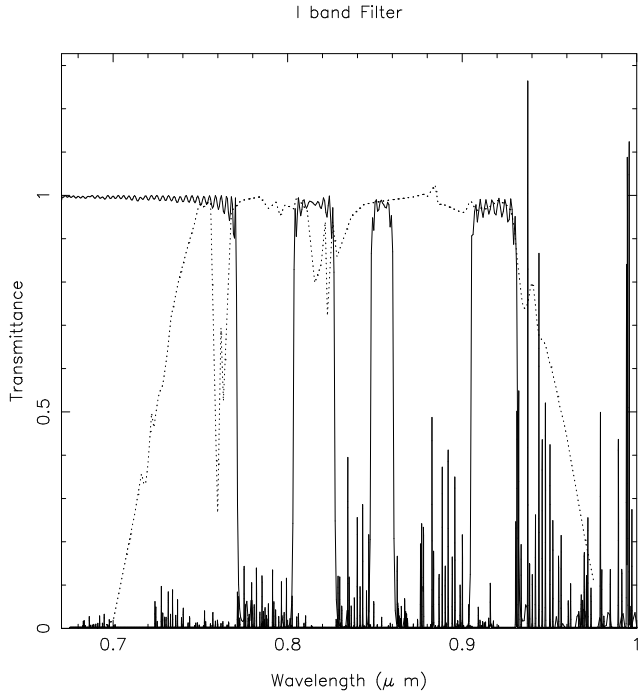} }
\caption{The transmittance profile as a function of wavelength for an
  I band OH suppression filter. The dotted line is a product of the
  atmospheric absorption and the assumed I band profile.}
\end{figure} 

Figure~7 shows the $I$ band equivalent of filter Jc. As for the $J$ band
filters, the transmittance profile has been calculated assuming a
collimated beam at normal incidence, and assuming non-dispersive
materials. A substrate refractive index of 1.47 was assumed and the
initial refractive index profile, expressed as a sum of ten
sinusoids, varied between 1.30 and 1.71. This was truncated at
1.35 to 1.62 before the  final optimisation. The properties of 
the filter are summarised in table~2.

The improvement in signal to noise ratio was calculated for a
magnitude 24 continuum object, and the results are summarised in
table~3. The corresponding values of the signal to noise ratio for a
600s integration time and a magnitude 24 1\arcsec\ continuum object
are 9.2 with no OH suppression and 10.6 with the OH suppression
filter. As can be seen, the results are much less spectacular then for
the $J$ band, although a small improvement in signal to noise ratio of
a factor of 1.16 can still be found. Greater improvements in signal to
noise ratio are possible but with increasingly complex transmission
profiles. The relatively poor performance is partly a result of the
distribution of line emission in the band but largely a result of the
relatively high fraction of the background intensity in the continuum.
Note that the improvement in the signal to noise ratio depends on the
balance between the background line and continuum intensities. The
values chosen here are conservative.  If the continuum is completely
neglected than the improvement in signal to noise ratio for a
magnitude 24 1\arcsec\ continuum object rises to 1.39, suggesting a
potential 39\% improvement in the signal to noise ratio if the
continuum is very small compared with the intensity of the emission
lines.

\section{Discussion}

\subsection{Filter Manufacture} 
The major challenge of the rugate technology has been to develop
design software compatible with the control systems of vacuum
deposition chambers. Coating designs which can be expressed as the sum
of up to a dozen sine waves appear to be the most straightforward to
manufacture. But it should be stressed that the technology is
sufficiently new that filter transmission profiles as complex as those
proposed in Fig. 2 have not been attempted by commercial companies
before. A constraint imposed at the outset was to limit the optical
coating thickness to $\approx 20\mu$m, and infrared coatings to
$\approx 40\mu$m, since thicker coatings are a major technical
challenge. Even these limits are at least twice the thickness of
routine multilayer coatings used in interference filters. At the
outset, the number of layers, and hence the coating thickness, relates
almost directly to how long the filter resides within the vacuum
chamber. Interference filters are produced with real-time quality
control and it becomes more difficult to control the transmissive
behaviour as more layers are laid down.

Motivated by our designs, B. Bovard has demonstrated with a
combination of tantalum pentoxide ($n_o = 1.5$) and silicon dioxide
($n_o = 2.3$) that the desired filter response in Fig. 2$ii$ can be
achieved with a 40$\mu$m thickness rugate coating. The narrowest rejection
band at 1170nm drives the total coating thickness; the broad rejection
bands are much less challenging.
For the Ta$_2$O$_5$:SiO$_2$ refractory oxide combination, the troughs
achieve blocking to better than 0.1\% while the peaks have steep sides
and average better than 90\% transmission. He finds the central
wavelength of each rejection band can be achieved to within 1\%, but
the width is more uncertain by as much as 5\% of the rejection band.
Bovard has achieved optical quality and a uniform spectral response
over a 150mm diameter and suggests that a 200mm diameter is within
reach.  Rugate coatings require intricate (mostly proprietary)
deposition techniques carried out with great care over several days in
a vacuum chamber. This makes them expensive with costs per filter
close to $\sim$US\$30,000, although possible approaches to reducing
costs include subdividing a coated blank with a diamond pencil.

It has been shown that effective $J$ band OH suppression filters can
be designed in principle but the robustness of such a filter is an
important consideration in practice. There are several effects which
serve to degrade filter performance. Thermal effects induce a bandpass
shift at the level of 0.2\AA\ K$^{-1}$. Small corrections are possible
with filter tilts if the original filter is specified at a temperature
warmer than under observing conditions. Deviations from optical
flatness are caused by microstructure within the coatings and by
bowing in the filter substrate due to strong surface tension (stress)
within the coating dielectric \cite{ennos,martin}.  Basic approaches
to reduce bowing are to apply the coating layers to both sides of
the substrate, or to bind the multilayer coating to a thick
($\sim$10mm) substrate, but a large optical thickness can seriously
degrade the optical performance of the camera. An alternative route is
to match materials with tensile and compressive
stress (e.g. zinc sulphide:cryolite, zinc sulphide:thorium fluoride).
This contrasts with Bovard's hard coating where both dielectrics are
compressive, a combination which has proven stability and longevity.

Small differences in manufacture induce in situ stress across the
plates, although ion bombardment can remedy this to some degree
\cite{martin}.  Deviations from optical flatness cause the filter
peaks to spread, thereby filling in the troughs.

The performance of the filter when expressed as an improvement in the
signal to noise ratio has a highly complicated dependence on the
transmittance profile. Some insight into the sensitivity of the signal
to noise improvement to the degradation of the filter profile can be
gained by fitting Butterworth functions to the low transmittance
notches in the `ideal' $J$ filter profile shown in figure~$2i$. The
Butterworth profile is given by:
\begin{equation}
\label{butterw}
f_n(\lambda) = \left({ 1 + \left({ \lambda - \lambda_c\over
        h/2}\right)^{2n} }\right)
\end{equation}
where $\lambda_c$ is the central wavelength of the notch and $h$ is
the FWHM (full width at half max) of the function. Figure~8 shows
three Butterworth functions with the same FWHM and central wavelength
as the square notches in figure~$2i$ but with $n=3, 6, 10$ and $30$. A
square profile results when $n\to \infty$. By varying $n$ and $h$ the
Butterworth functions can be used to model the effects of both changes
in the width of the notches and increasing roundness of the notches.
Figure~9 shows the variation in the improvement in signal to noise
ratio as a function of incremental change in the half width of the
three notches in figure~8 for $n=\infty$ (square wave profile), and
$n=3, 6, 10$ and $30$.  A degradation of the profile through small
irregularities would both smooth the profile (ie decrease $n$) and
change the half width so extrapolation to a real degraded filter is
difficult. However, it can be seen that the signal to noise improvement
is more tolerant to a broadening of the notches (increase in
halfwidths) than to a narrowing of the notches. The percentage
background intensity retained can increase rapidly as the notch
shrinks and additional strong OH lines are admitted.  The percentage
drop in object flux resulting from a broader notch is much smaller
than the percentage rise in background resulting from a narrow notch.
 
\begin{figure}
\label{f-butter}
\ \centering { \epsfxsize=3.3in \epsfbox{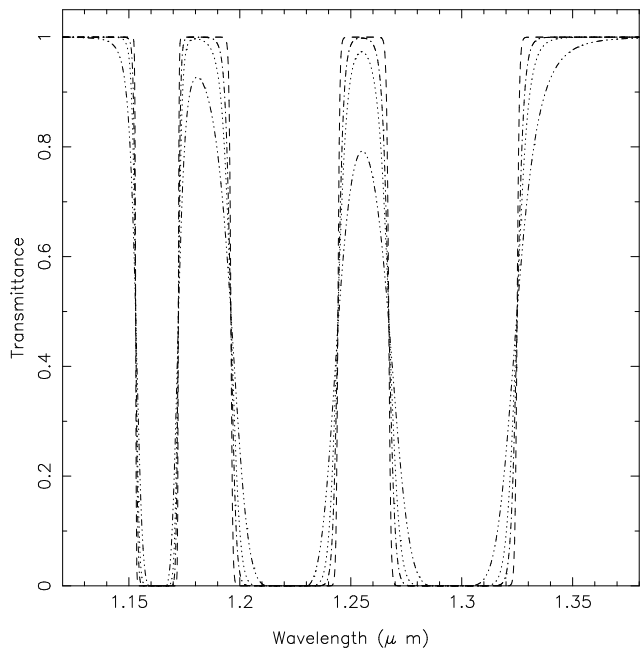} }
\caption{ Butterworth profiles with the same central wavelengths and
  half widths as the notches in the `ideal' $J$ filter but with $n=30$
  (dashed line), $n=10$ (dot-dash), $n=6$ (dotted)  and $n=3$
  (dash-dot-dot-dot).}
\end{figure}

\begin{figure}
\label{f-perturb}
\centering { \epsfxsize=3.3in \epsfbox{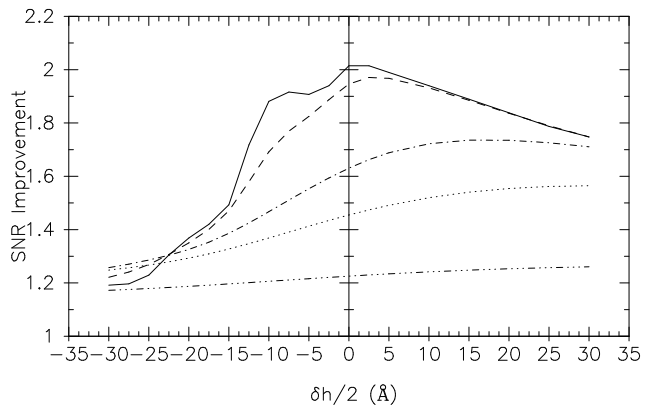} }
\caption{The variation of signal to noise improvement as a function
  of the incremental change in the half widths of Butterworth profiles
  with $n=\infty$ (solid line), $n=30$ (dashed line), $n=10$
  (dot-dash), $n=6$ (dotted line) and $n=3$ (dash-dot-dot-dot). When
  $\delta h=0$ these correspond to the transmittance profiles shown in
  figure 8. Positive $\delta h$ corresponds to broader low transmittance
  notches.}
\end{figure}

\subsection{Instrumental Applications}

Up to this point, our calculations have assumed a collimated beam at
normal incidence. However, wide field imaging surveys are carried out
with fast cameras in which blocking filters are traditionally used in
the converging beam. It is well known that such beams modify both the
first and second moments of the filter transmission profile. The
former is rectified by shifting the desired filter response to red
wavelengths at the design stage. The latter limits the speed of the
beam depending on the precise filter structure and the angle subtended
by the detector seen from the camera lens. Analytic approximations to
the first four moments are given by Bland-Hawthorn
\shortcite{hawthorn}. We find that the rugate profiles presented in
Fig. 2 suffer negligible degradation in beams slower than f/3, inclusive
of non-telecentric rays over a 20\arcmin\ field.

\begin{figure} 
\label{f_ijfilter} 
  \ \centering { \epsfxsize=3.3in \epsfbox{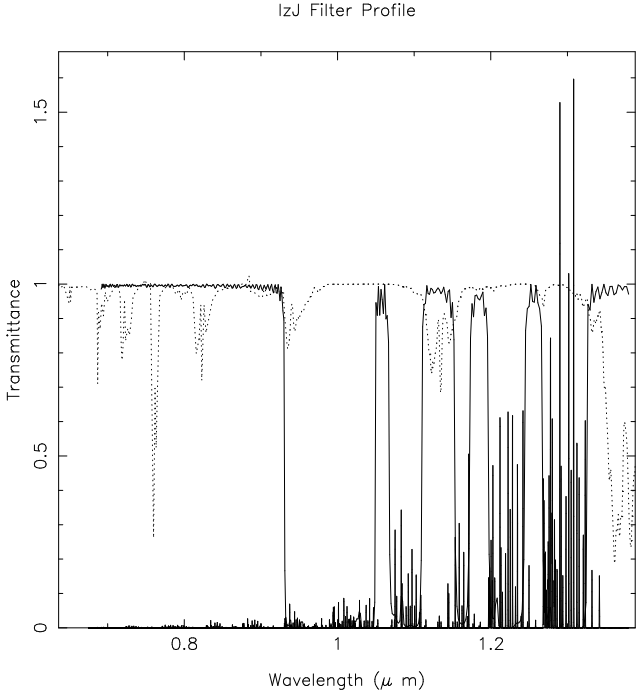} }
\caption{A transmittance profile as a function of wavelength for
an $IzJ$ filter that would reduce the sky background in $J$ to the
same level as in $I$. The dotted line is the atmospheric transmittance.}
\end{figure} 
  
The new-generation Rockwell HAWAII arrays provide high (60\%)
quantum efficiency in the near-infrared with moderate performance
(50\%) in the $I$ and $z$ bands. An
optical system built around a HAWAII array could achieve comparable
sensitivity in standard $I$ and rugate $J$ bands. Such a device
could also capitalize on an $IzJ$ filter  to enhance the 
sensitivity at near infrared wavelengths compared to conventional 
optical bands. For example, figure~10 illustrates a
transmission profile that will reduce the sky background for 
$\lambda >  1$\hbox{$\mu\rm m$} to the same level as the sky background for
$\lambda < 1$\hbox{$\mu\rm m$}, thereby allowing a relatively 
uniform $IzJ$ band to be defined. Clearly this profile also incorporates
an extended $J$ or $zJ$ filter from $1.05$ to $1.38{\mu\rm m}$
which gives access to an additional clean spectral window not used in
the $J$ filters discussed earlier.

A very different application of rugate technology would be to devise a
beam splitter whose complementary images amplify some characteristic
in the spectrum of a class of objects. This idea, discussed in the
next section, has the potential to usher in a spate of inexpensive
astronomical imagers as the technique can be exploited with a
rudimentary optical design.

\subsection{Scientific Applications}

Rugate filters are expected to increase observing efficiency at most
sites, and therefore the scientific applications are numerous.  One
example, discussed further below, is very low mass (VLM) stars and brown
dwarfs. These are so rare and faint that they require dedicated
surveys over huge volumes -- which means either very wide field or
very deep surveys. The most reliable survey diagnostic is the $I-J$
colour \cite{leggett}, which is exactly the wavelength range where
rugate technology will produce the largest gain in sensitivity.

For continuum sources, we have seen that signal:noise gains as high as
2 are possible in the rugate $J$ band. This may find important uses in
studies of galaxy evolution since the K-corrections at $J$ are much
closer to those at $K$ (Fig.~11) than to optical bands
\cite{glazebrook}.  In particular, the  K-corrections for the different
morphological classes are remarkably uniform for rugate filters $Jc$
and $Jd$ over the redshift interval $z=1-3$. Thus, evolutionary
signatures should be much easier to monitor in a rugate $J$ filter
than at $I$, and the higher sensitivity at $J$ compared to $K$ makes
it the preferred near-infrared band out to at least $z=2$. The
morphological types start to diverge when the H,K break enters the
band: this occurs at $z=1$ for $I$ and $z=2$ for $J$. The
K-corrections for the rugate filters are remarkably stable as they
agree to 0.1 mag out to $z=3$.

\begin{figure}
\label{f-kcorr}
\ \centering { \epsfxsize=3.3in \epsfbox{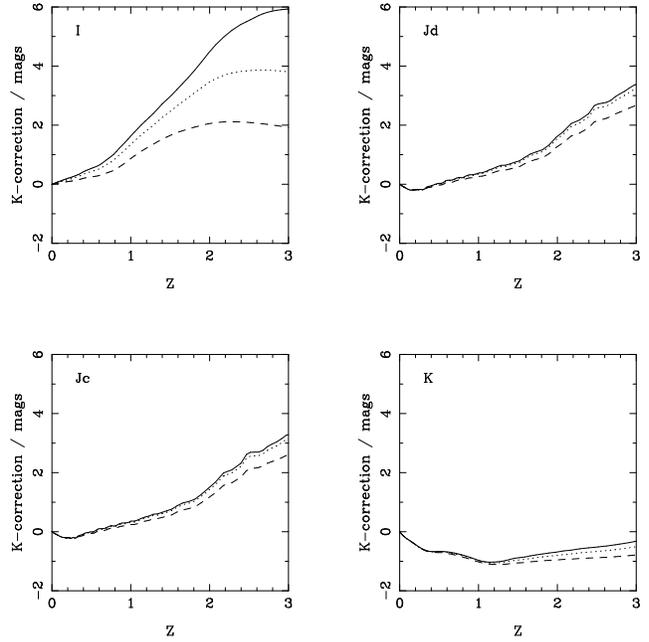} }
\caption{
  K-corrections at $I$ and $K$ compared to two rugate filter designs,
  $Jc$ and $Jd$ (see Fig.2).   The galaxy types shown are
  elliptical galaxies (solid lines), early-type spirals (dotted
  lines), and late-type/irregular spirals (dashed lines).  The small
  bumps on the rugate curves are due to the H,K break moving through
  the sub-bands.}  
\end{figure}

\begin{table*} 
\begin{minipage}{140mm}
  \centering
\caption{Rugate filter $Jc$: transmitted redshift ranges for prominent emission lines}
\label{Jc-z} 
\vspace{0.3cm}
\begin{tabular}{ccc} 
  \hline
H$\alpha$ &  [OIII]5007 &   [OII]3727\\
  \hline
0.71-0.76 & 1.24-1.30 & 2.01-2.09\\
0.79-0.82 & 1.34-1.39 & 2.14-2.21\\
0.90-0.93 & 1.49-1.53 & 2.34-2.40\\
1.02-1.09 & 1.65-1.74 & 2.56-2.68\\
  \hline
\end{tabular}
\end{minipage}
\end{table*}

Even larger gains are possible in observations of emission-line
sources, assuming that the spectral line falls within one of the
transmitting windows. If the object signal arises in a discrete,
narrow interval transmitted by the filter, no signal from the object
is lost by the OH suppressing bands, in which case signal:noise gains
as high as 4 are possible.  In redshift surveys, roughly half of the
survey volume is lost to the suppressing bands, although this
overstates the problem if the line-emitting source transmits in
several bright lines within the broad window. To date, searches for
high redshift galaxies have concentrated on only a few strong lines
from Ly$\alpha$ to H$\alpha$. In Table~\ref{Jc-z}, we show the
transmitted redshift ranges for three prominent emission lines. The
redshift slices transmitted by rugate filter $Jc$ are well matched to
the redshift range over which the star formation rate density is
expected to peak in the early universe \cite{ellis}.

We reiterate a point made by Jones et al.  \shortcite{jones}
that in circumstances where a rugate filter does not produce a
substantial gain, this may still be the preferred option due to the
longer exposure times and suppressed background variation between
exposures. It is anticipated that this will lead to much better
image mosaics than have been possible to date.

This paper has concentrated on the use of rugate filters for OH
suppression but the great versatility of rugate technology could find
many uses within astronomy. For example, in the case of brown dwarfs
the ability to define irregular bands within a single filter suggests
an alternative approach to finding these elusive objects. At
near-infrared wavelengths, the spectrum of Gliese 229B \cite{geballe}
has a roughly sawtooth appearance. A rugate filter could be made to
operate as a beam splitter such that the peaks of the sawtooth are
transmitted and the troughs are reflected by the filter.  If the
separate paths are imaged onto two detectors, candidate objects would
appear bright in one image and dark at the same location on the other
detector. The graded index coatings are sufficiently flexible that the
precise positions of the peaks and troughs could be chosen to share
the sky background equally between both filters. Thus, rugate
technology could give rise to a range of niche instruments designed to
find specific classes of astrophysical objects with distinctive
spectra.

In summary, interference filters offer a practical approach to
achieving significant gains in OH-suppressed imaging. We recognise
that more sophisticated techniques could be conceived. 
One possibility is the use of an OH absorption cell to 
absorb the OH line emission directly. But the producion and 
maintainance of a vibrationally excited OH population with a sufficiently 
high column density would be a major technical challenge. The 
vibrational bands contributing in the $J$ band
are the $\nu=8\to 5$, $7\to 4$ and $6 \to 3$ bands, thus OH
of vibrational levels at least up to $\nu=5$ would be required
at relatively high concentrations.   Relaxation of
the vibrational states would rapidly depopulate the higher
vibrational levels and some means of maintaining the 
vibrationally excited population would be required \cite{kond}. This simple 
idea is likely to prove complex and costly to implement. Although less 
elegant, the OH suppression filters can be produced relatively 
cheaply with existing technologies and could easily be incorporated
into conventional instruments.

\section{Summary} 

The sensitivity of astronomical observations in the near infrared is
limited by the presence of intense atmospheric OH emission bands. The
conventional approach to improving sensitivity has been to increase
the telescope aperture. However, in limited cases, noise reduction can
achieve major gains in sensitivity. In this paper we have shown that
the simulated rugate filters can achieve extraordinary gains for
imaging at $J$, and modest gains through $I$ and $z$, at the expense of
loss in bandpass. A similar $H$ band filter would require a much more
complicated transmission profile and, since the transmission profiles
proposed for the $I$ and $J$ bands are already near the limit of
currently realisable profiles, it was not considered in detail. Plans
are already well advanced to manufacture a rugate $J$ filter for wide
field imaging at the Anglo-Australian 3.9m Telescope.  We have
provided  examples of astrophysical studies that will clearly
benefit from such a filter, as we propose to demonstrate in subsequent
papers.

\section{Acknowledgements}
We are indebted to K. Glazebrook for producing Fig.~11 and
for helpful dialogues throughout the work.  We would also like to thank
P.A. Markham, I.J. Lewis and C.G. Tinney for useful comments. On the
technical aspects, R.P. Netterfield and B.G.  Bovard offered
constructive advice at critical stages of the work.  We acknowledge
the stimulating atmosphere at the Anglo-Australian Observatory which
has sustained much of the development work described here.

%bibliography 

\begin{appendix}
\section{Genetic Algorithms} 
 
Genetic algorithms are optimisation techniques that draw on the
Darwinian idea of evolution through survival of the fittest
\cite{gabook}. They are particularly suited to non-linear optimisation
(minimisation or maximisation) problems with many distinct local
minima where gradient following optimisation techniques will fail to
find the global minimum.  There are a number of possible
implementations of the method. The method used here closely follows
the algorithm described by Charbonneau \shortcite{ga}.

The basic steps are as follows:
\begin{enumerate}
\item{Form a random initial population of solutions}
\item{Assign a fitness value to each member of the population related
    to how closely it approached some desired solution}
\item{Choose pairs of individuals from the population and form
    `offspring' by breeding the two individuals. Introduce mutations
    into the genes.}
\item{Establish the fitness of the offspring and, if they are fitter
    than the weakest member, insert them into the population}
\item{Repeat 2--4 until convergence is reached}
\end{enumerate}
   
An initial population of $N_p$ individuals is created randomly. In the
present case each member of the population corresponds to a complete
set of filter parameters, $\delta_i, \alpha_i, \phi_i$ for $i=1,m$
where $m$ is the number of sinusoids in equation~\ref{eq_riform}.
These \em phenotypes \em are decoded versions of their associated \em
genotypes \em, the set of `genes' that determine the parameters. Each
individual has associated with it $3m$ genes, each gene being a
sequence of 5 integers between 0 and 9. The associated parameters are
then given by interpreting the 5 integers as the first five
significant figures of a real number between $0.0$ and $0.99999$. The
parameters are then given by:
\begin{equation}
  \delta_i = \delta_{max} \times x_{3i-2}
  \hbox{\hspace{1cm}(i=1,2,3,...,m)}
\end{equation}
\begin{equation}
  \alpha_i = (\alpha_{max} - \alpha_{min})\times x_{3i-1} +
  \alpha_{min} \hbox{\hspace{1cm}(i=1,2,3,...,m)}
\end{equation}
\begin{equation}
  \phi_i = 2\pi \times x_{3i} \hbox{\hspace{1cm}(i=1,2,3,...,m)}
\end{equation}
where $\alpha_{max},\alpha_{min}$ and $\delta_{max}$ are imposed
constraints on the allowed range of the parameters. Here the $x_k$ are
given by:
\begin{equation}
  x_k = \sum_{j=1}^5 10^{-j} I_{5(k-1)+j}
\end{equation}
where the $I_j$ are the 5 integers representing the individual gene.
\begin{equation}
  g_k = (I_{5k-4},I_{5k-3},I_{5k-2},I_{5k-1},I_{5k})
  \hbox{\hspace{1cm}(k=1,3m)}
\end{equation}
The $x_k$ serve to translate between the $5\times 3m$ integers
representing the genotype, and the physical properties $\delta_i$,
$\alpha_i$ and $\phi_i$.

The population is ranked in order of fitness, however fitness is
defined.  In the present case the fitness is defined either as being
related to the least squares deviation between the filter
transmittance profile (Equation~\ref{eq_fit1}) or as an explicit
function of the improvement in signal:noise ratio (Eq.~\ref{eq_fit2}).

Once the population is established, breeding can begin. Individuals
are chosen for breeding with a probability proportional to their
fitness ranking \cite{ga}. When a pair of individuals are chosen the
breeding operation is initialised by making copies of the parents. For
example:
\begin{equation}
  {\hbox{Parent 1} = (I_1,I_2,I_3,{\dots},I_{5\times 3m}) \nonumber }
\end{equation}
\begin{equation}
  {\hbox{Parent 2} = (J_1,J_2,J_3,{\dots},J_{5\times 3m})}
\end{equation}
On breeding, and exchange of genes takes place.  A random number is
generated between 0 and 1. If this is less than a preset `cross-over'
probability then a random point is selected in the two genotype arrays
of $5\times 3m$ integers, say position $n$. From that point forward
the genes of the two parents are crossed to create two new
individuals:
\begin{equation}
  {\hbox{Offspring 1} =
    (I_1,I_2,I_3,{\dots},I_{n-1},J_n,J_{n+1},{\dots},J_{5\times 3m})
    \nonumber }
\end{equation}
\begin{equation}
  {\hbox{Offspring 2} =
    (J_1,J_2,J_3,{\dots},J_{n-1},I_n,I_{n+1},{\dots},I_{5\times 3m}) }
\end{equation}
Mutation may also occur during breeding. A random number is selected
for each of the $5\times 3m$ integers, if this is less than the
(small) mutation probability the integer is replaced by a randomly
chosen digit.

The genotypes of the two offspring are translated back into two sets
of filter parameters and the fitness of the individuals is calculated.
The population replacement strategy used is the `replace worst'
strategy described by Charbonneau \shortcite{ga}. If the fitness of
the new individuals was greater than the weakest members of the
population the weakest are replaced by the new individuals. A
generation is defined as a series of $N_p$ breeding operations.

The rate of convergence of the solutions is rapid initially but
becomes slow as the optimal solution is approached. The rate of
convergence can be controlled to some extent by varying the mutation
rate but other more direct optimisation techniques may be necessary to
fine tune the solutions.
\end{appendix} 
 
\label{lastpage} 


\begin{thebibliography}{1} 
%\bibitem[\protect\citename{ }19 ] {}
\bibitem[\protect\citename{Abrams et al.  }1994] {abrams}
  Abrams M.C., Davis S.P., Rao M.L.P., Engleman R.Jr., Brault J.W.,
  1994, ApJS, 93, 351
\bibitem[\protect\citename{Bates \& Nicolet }1950]{chemistry} 
  Bates D.R. \& Nicolet M., 1950, JGR, 55, 301
\bibitem[\protect\citename{Bates }1982]{bates} Bates D.R., 1982, in
  Massey H.S.W.\& Bates D.R., eds, Atmospheric Physics and
  Chemistry v1, Applied Atomic Collision Physics, 149
\bibitem[\protect\citename{Bland-Hawthorn }1997]{hawthorn}
  Bland-Hawthorn J. 1997, PASP, submitted
\bibitem[\protect\citename{Broadfoot \& Kendall }1968] {broadfoot}
  Broadfoot A.L. \& Kendall K.R., 1968, JGR, 73 426
\bibitem[\protect\citename{Charbonneau }1995]{ga} 
  Charbonneau P. ,  1995, ApJS, 101, 309 
\bibitem[\protect\citename{Content }1996] {content2} 
  Content R. 1996, ApJ, 464, 412 
\bibitem[\protect\citename{Content \& Angel }1994] {content1} 
    Content R. \& Angel R., 1994, in Crawford D.L. and Craine E.R.,
    eds,  Proc. SPIE 2198, Instrumentation in Astronomy VIII, 757
\bibitem[\protect\citename{Dobrowolski \& Verly }1993] {opticsrefs} 
  Dobrowolski, J.A. \& Verly, P.G., eds, 1993,
  Proc.SPIE. 2046, Inhomogeneous and Quasi Inhomogeneous Optical Coatings  
\bibitem[\protect\citename{Ellis }1997]{ellis}
  Ellis R.S. 1997, Annual Reviews, 35, in press (astro-ph/9704019)
\bibitem[\protect\citename{Ennos }1966]{ennos}
  Ennos A.E. 1966, Appl.Opt. 5, 51
\bibitem[\protect\citename{Fabricus }1992]{quintics2} 
  Fabricus H.  1992 Appl.Opt. 31 5191 
\bibitem[\protect\citename{Fukugita et al. }1996] {fuku}
  Fukugita M., Ichikawa T., Gunn J.E., Doi M., Shimasaku K. \&
  Schneider D.P., 1996, AJ, 111, 1748
\bibitem[\protect\citename{Geballe et al. }1996]{geballe} 
  Geballe T.R., Kulkarni S.R., Woodward C.E. \&
  Sloan, G.C.  1996, ApJ, 467, 101
\bibitem[\protect\citename{Glazebrook }1997]{glazebrook}
    Glazebrook K., 1997, in  Mamon G.A., Thuan T.X. \& Van J.T.T., eds,
    Extragalactic Astronomy in the Infrared, Editions Frontieres
\bibitem[\protect\citename{Goldberg }1989]{gabook} 
   Goldberg D.E., 1989,  Genetic Algorithms, Addison-Wesley 
\bibitem[\protect\citename{Greiner }1996]{greiner}
  Greiner H. , 1996, Appl.Opt., 35, 5477
\bibitem[\protect\citename{Harrison \& Kendall }1973] {harrison}
  Harrison \& Kendall, 1973, Planet.Space.Sci., 21, 1731
\bibitem[\protect\citename{Herbst }1994]{herbst} 
  Herbst T.M. , 1994, PASP, 106, 1298 
\bibitem[\protect\citename{Iwamuro et al.  }1994]
  {iwamuro} Iwamuro F., Maihara T., Oya S., Tsukamoto H., Hall D.N.B.,
  Cowie L.L., Tokunaga A.T. \& Pickles A.J., 1994, PASJ, 46, 515
\bibitem[\protect\citename{Johnson \& Crane }1993]{johnson} 
  Johnson W.E. \& Crane R.L., 1993, in Dobrowolski, J.A. \& Verly, P.G.,
  eds,  Proc.SPIE.2046, Inhomogeneous and Quasi Inhomogeneous 
  Optical Coatings, 88
\bibitem[\protect\citename{Johnston \& Broadfoot }1993] {johnston}
  Johnston J.E. \& Broadfoot A.L., 1993, JGR, 98(A12),
  21593 
\bibitem[\protect\citename{Jones et al.  }1997]{jones}
  Jones D.H., Bland-Hawthorn J. \& Burton M.G.,
  1996, PASP, 108, 929 
\bibitem[\protect\citename{Kondratyev  }1969] {kond}
  Kondratyev K.Ya., 1969,   Radiation in the Atmosphere, Academic Press,
  New York and London
\bibitem[\protect\citename{Koorneef }1983]{koorneef} 
  Koorneef J. 1983, A\&A, 128, 84 
\bibitem[\protect\citename{Leggett }1992]{leggett}
  Leggett, S.K. 1992, ApJS, 82, 351 
\bibitem[\protect\citename{Le Texier et al.  }1987] {letexier} 
  Le Texier H., Solomon S. \& Garcia R.R., 1987, Planet.Space.Sci.  35, 977
\bibitem[\protect\citename{Levi }1980]{levi} 
    Levi L., 1980,  
    Applied Optics: A Guide to Optical System Design  v2, Wiley
  Interscience, chapter 11 
\bibitem[\protect\citename{Llewellyn et al. }1978]{llewellyn} 
    Llewellyn E.J., Long B.H. \& Solheim B.H.,
    1978, Planet.Space.Sci., 26, 525 
\bibitem[\protect\citename{Maihara et al.  }1993] {maihara2} 
  Maihara T., Iwamuro F., Yamashita T.,
  Hall D.N.B., Cowie L.L., Tokunaga A.T., \& Pickles A., 1993, PASP,
  105, 940 
\bibitem[\protect\citename{Maihara }1994] {maihara1}
  Maihara T., Iwamuro F., Oya  S., Tsukamoto H., Hall D.N.B., Cowie L.L.,
  Tokunaga A.T., \& Pickles A.J., 1994, in Crawford D.L. and Craine E.R.,
    eds,  Proc. SPIE 2198, Instrumentation in Astronomy VIII, 194
\bibitem[\protect\citename{Martin et al.  }1991]{martin} 
  Martin P.J., Netterfield R.P., Kinder T.J. \& Stanbouli V.  1991, 
  Ap.Phys.Lett., 58: 22, 2497 
\bibitem[\protect\citename{Martin et al. }1995]{ga_opt} 
  Martin S., Rivory J. \& Schoenauer M. , 1995,
  Appl.Opt., 34, 2247 
\bibitem[\protect\citename{Noxon }1978]{noxon}
  Noxon J.F., 1978, Planet.Space.Sci., 26, 191
\bibitem[\protect\citename{Oliva \& Origla }1992] {oliva} 
  Oliva E. \& Origla L., 1992, A\&A, 254 466
\bibitem[\protect\citename{Osterbrock et al.  }1997] {osterbrock}
  Osterbrock D.E., Fulbright J.P., Bida T.A., 1997, PASP, 109, 614
\bibitem[\protect\citename{Pich\'e  }1996]{piche2}
  Pich\'e F., 1996, Private Communication
\bibitem[\protect\citename{Pich\'e et al. }1997]{piche} 
  Pich\'e F., Parry I.R., Enrico K., Ellis R.S.,
  Pritchard J., Mackay C.D. \& McMahon R.G., 1997, 
  in Ardeberd A., ed, Proc. SPIE 2871,  Optical
    Telescopes of Today and Tomorrow: Following in the Direction of
    Tycho Brahe,  1332
\bibitem[\protect\citename{Ramsay et al.  }1992] {ramsay} 
  Ramsay S.K., Mountain C.M. \& Geballe T.R., 1992, MNRAS, 259, 751
\bibitem[\protect\citename{Sobolev }1978]{sobolev} 
  Sobolev V.G., 1978, Planet.Space.Sci., 26, 703 
\bibitem[\protect\citename{Southwell }1985]{southwell} 
Southwell W.H. 1985, Appl.Opt., 24, 456
\bibitem[\protect\citename{Southwell \& Hall }1989]{quintics1}
  Southwell W.H. \& Hall R.L. 1989 Appl.Opt. 28 2949
\bibitem[\protect\citename{Southwell }1989]{apodise} 
  Southwell W.H. 1989 Appl.Opt. 28 5091 
\bibitem[\protect\citename{Steed \& Baker }1978] {steed} 
  Steed A.J. \& Baker D.J., 1978, Appl.Opt., 18, 3386
 

\end{thebibliography}
\end{document}